\documentclass[global,twocolumn]{svjour}
\usepackage{latexsym}
\usepackage{graphicx}
\journalname{myjournal}
\begin{document}

apb-fissionfusion15.dvi keepcomments
\title{Introducing the Fission-Fusion Reaction Process:
       Using a Laser-Accelerated Th Beam to produce Neutron-Rich Nuclei
       towards the $N=126$ Waiting Point of the $r$-Process}

\author{D. Habs\inst{1,2}, P.G. Thirolf\inst{1}, M.~Gross\inst{1},
        K. Allinger\inst{1}, J. Bin\inst{1}, A. Henig\inst{1},
        D. Kiefer\inst{1}, W. Ma\inst{2} \and J. Schreiber\inst{2}
}                     
\offprints{}          
\institute{$^1$ Fakult\"at f\"ur Physik, Ludwig-Maximilians Universit\"at M\"unchen,
                D-85748 Garching, Germany \\
           $^2$ Max-Planck-Institut f\"ur Quantenoptik, D-85748 Garching, Germany}
\date{Received: date / Revised version: date}

\titlerunning{The fission-fusion reaction process}
\maketitle

\begin{abstract}

We propose to produce neutron-rich nuclei in the range of the astrophysical
$r$-process (the rapid neutron-capture process) around the waiting point 
$N=126$~\cite{kratza07,arnould07,janka09} by fissioning a dense 
laser-accelerated thorium ion bunch in a 
thorium target (covered by a polyethylene layer, CH$_2$), where the light 
fission fragments of the beam fuse with the light fission fragments of the 
target. 
Via the 'hole-boring' (HB) mode of laser Radiation Pressure Acceleration 
(RPA)~\cite{robinson09,henig09,tajima09} using
a high-intensity, short pulse laser, 
very efficiently bunches of $^{232}$Th with solid-state density can be generated
from a Th layer (ca. 560~nm thick), placed beneath a deuterated polyethylene
foil (CD$_2$ with ca. 520~nm), both forming the production target. 
Th ions laser-accelerated to about 7 MeV/u will pass through 
a thin CH$_2$ layer placed in front of a thicker second 
Th foil (both forming the reaction target) closely behind the production target 
and disintegrate into light and heavy fission fragments.  
In addition, light ions (d,C) from the CD$_2$ production target
will be accelerated as well to about 7 MeV/u, inducing the fission process 
of $^{232}$Th also in the second Th layer. 
The laser-accelerated ion bunches with solid-state density,
which are about $10^{14}$ times more dense than classically accelerated 
ion bunches, allow for a high probability that generated fission products
can fuse again when the fragments from the thorium beam strike the 
Th layer of the reaction target.  \\
In contrast to classical radioactive beam facilities, where intense but 
low-density radioactive beams of one ion species are merged with stable targets, 
the novel fission-fusion process draws on the fusion between 
neutron-rich, short-lived, light fission fragments both from beam and target.
Moreover, the high ion beam density may lead to a strong collective
modification of the stopping power in the target by 'snowplough-like'
removal of target electrons, leading to significant range enhancement,
thus allowing to use rather thick targets.\\
Using a high-intensity laser with two beams with a total energy of 300 J, 32~fs 
pulse length and 3 $\mu$m focal diameter, as, e.g., envisaged 
for the ELI-Nuclear Physics project 
in Bucharest (ELI-NP)~\cite{eli-np}, order-of-ma\-gni\-tude estimates 
promise a fusion yield of about $10^3$ ions per laser pulse in the mass range 
of $A=180-190$, thus enabling to approach the $r$-process waiting point at N$=$126.
First studies on ion acceleration, collective modifications of the stopping behaviour
and the production of neutron-rich nuclei can also be performed at the 
upcoming new laser facility CALA (Center for Advanced Laser Applications) in 
Garching.

\end{abstract}

\section{Introduction}
  \label{intro}

Elements like platinum, gold, thorium and uranium are produced
via the rapid neutron capture process ($r$-process) at astrophysical 
sites like merging neutron star binaries or (core collapse) supernova 
type II explosions
with outbursts of very high neutron density in the range of $10^{21}-
10^{30}/cm^3$. We aim at improving our understanding of these nuclear processes
by measuring the properties of heavy nuclei on (or near) the $r$-process path.
According to a recent report by the US National Research Council of the National
Academy of Science, the origin of the heaviest elements remains one of
the 11 greatest unanswered questions of modern physics \cite{haseltine02}.
While the lower path of the $r$-process for the production of heavy elements 
is well explored, the nuclei around the $N=126$ waiting point critically 
determine this element production mechanism. At present basically 
nothing is known about these nuclei.

\begin{figure*}[]
\centerline{\includegraphics[width=.8\textwidth]{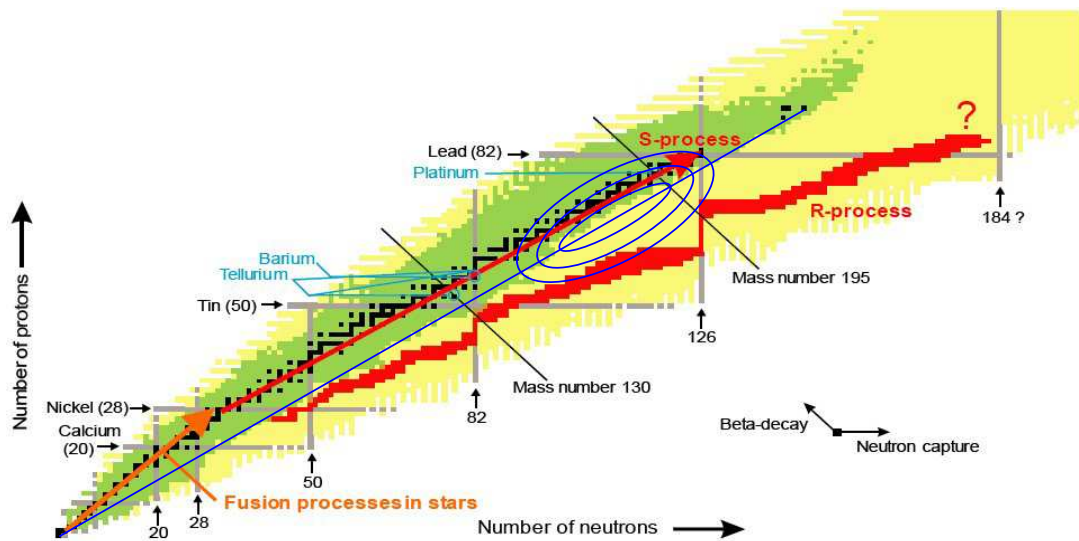}}
   \caption{Chart of the nuclides indicating various pathways for astrophysical 
            nucleosynthesis: thermonuclear fusion reactions in stars 
            (orange vector), $s$-process path (red vector) and the $r$-process 
            generating heavy nuclei in the Universe (red pathway). 
            The nuclei marked in black indicate 
            stable nuclei. For the green nuclei some nuclear properties are known, 
            while the yellow, yet unexplored regions extend to the neutron and proton 
            drip lines. The blue line connects nuclei with the same neutron/proton 
            ratio as for (almost) stable actinide nuclei. On this line the maximum 
            yield of nuclei produced via fission-fusion (without neutron evaporation) 
            will be located. The elliptical conture lines correspond to the expected 
            maximum fission-fusion cross sections decreased to 50\% ,10\% and 0.1\%,
            respectively, for primary $^{232}$Th beams.}
   \label{fig1}
\end{figure*}

Special ingredients of this proposal are: i) The very efficient
Radiation Pressure Acceleration (RPA) mechanism for laser-based ion
acceleration, especially exploiting the 'hole-boring' mode~\cite{robinson09} 
producing pancake-like beam bunches
of solid-state density. ii) The strongly reduced stopping power of these
dense bunches in a second thick Th target, where the decomposition into fission
fragments and the fusion of these fragments takes place.
After the laser flash we want to extract rather long-lived isotopes
($>$ 100 ms) in flight, separate them e.g. in a (gas-filled) recoil
separator and study them via decay spectroscopy or lifetime and
nuclear mass measurements.

In the following we outline the relevance of the project for nuclear
astrophysics, describe the new laser acceleration scheme and in particular
the new fission-fusion reaction method. Finally the planned ELI-Nuclear Physics
facility will be briefly introduced, where the production of these nuclei 
and the experiments to measure their properties will be realized. 


\section{The Relevance of the N=126 Waiting Point for Nuclear Astrophysics}
  \label{sec:astrophysics}

Fig.~\ref{fig1} shows the nuclidic chart marked with different 
nucleosynthesis pathways for the production 
of heavy elements in the Universe: the thermonuclear fusion processes in stars 
producing elements up to iron (orange arrow), 
the slow neutron capture process ($s$-process) along the valley of stability 
leading to about half of the heavier nuclei (red arrow)
and the rapid neutron capture process ($r$-process) proceeding along  
pathways with neutron separation energies $S_n$ in the range of 2--3~MeV. 
In this scenario, rather neutron-rich nuclei are populated in an intense 
neutron flux \cite{rolfs91}. The $r$-process path exhibits
characteristic vertical regions for constant magic neutron numbers 
of 50, 82 and 126, where the $r$-process is slowed down due to low neutron 
capture cross sections when going beyond the magic neutron numbers. 
These decisive bottlenecks of the $r$-process flow are called 
waiting points \cite{kratz07}. 

The astrophysical site of the $r$-process nucleosynthesis is still 
under debate: it may be cataclysmic core collapse supernovae (II) explosions 
with neutrino winds \cite{arnould07,janka09,janka07,thielemann04} or 
mergers of neutron-star binaries \cite{freiburghaus99,janka08,thielemann05}. 
The $r$-process element abundances from galactic halo stars tell us 
that the $r$-process site for lighter and heavier neutron capture processes may 
occur under different astrophysical conditions \cite{kratz07}.
For the heavier elements beyond barium, the isotopic abundancies are always 
very similar (called universality) and the process seems to be very robust. 
Perhaps also the recycling of fission fragments from the end of the $r$-process
strengthens this stability. Presently, it seems more likely that a 
merger of neutron star binaries is the source for the heavier $r$-process 
branch, while core collapsing supernova explosions
contribute to the lighter elements below barium. The modern nuclear
equations of state, neutrino interactions and recent supernova explosion 
simulations~\cite{janka09} lead to detailed discussions of the 
waiting point N$=$126. Here measured nuclear properties along the N$=$126 
waiting point may help to clarify the sites of the $r$-process.

\begin{figure}[t]
\centerline{\includegraphics[width=70mm]{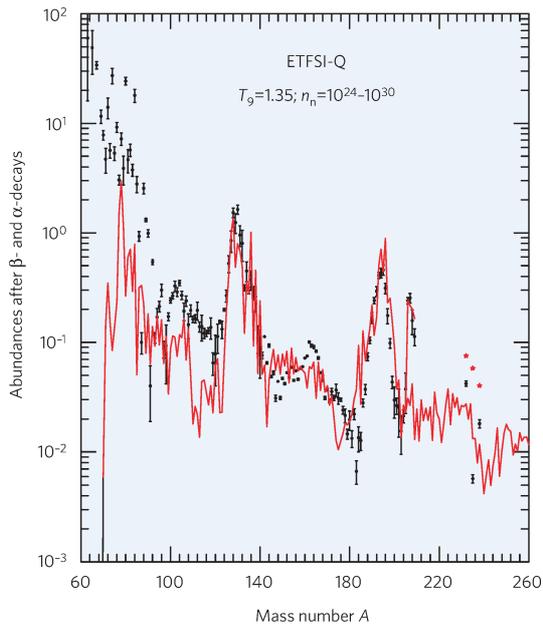}}
   \caption{Observed elemental solar abundances in the $r$-process 
            mass range (black symbols with error bars) in comparison with calculated 
            abundances (red line and symbols), normalized to silicon $= 10^6$. 
            The theoretical predictions show 
            the elemental abundances for stable isotopes after $\alpha$ 
            and $\beta$ decay as obtained in the ETFSI-Q mass 
            model~\cite{kratza07,pearson96} for a wide range of neutron 
            densities $n_n$ (in $1/cm^3$) and temperatures $T_9$ (in units 
            of $10^9$K) and including shell quenching effects. Included 
            with permission from~\cite{cowan}.}
   \label{fig2}
\end{figure}

Fig.~\ref{fig2} shows the measured solar elemental abundances of the 
$r$-process nuclei together with a calculation, where masses from 
the Extended Thomas-Fermi plus Strutinski Integral 
(ETFSI) mass model~\cite{pearson96} have 
been used together with several neutron flux components, characterized by 
a temperature $T_9$, neutron densities $n_n$ and expansion time scales. 
A quenching of shell effects~\cite{dobaczewki84} was assumed in the
nuclear mass calculations to achieve a better agreement between
observed and calculated abundances. 

The three pronounced peaks visible in the abundance distribution 
seem to be of different origin, which is also reflected in the 
theoretical calculations shown in Fig.~\ref{fig2}, where 
contributions from different temperatures and neutron densities are 
superimposed to the observed data. 
We note the pronounced third peak in the
abundance distribution around $A=180-200$, corresponding to the group 
of elements around gold, platinum and osmium, where until now no experimental 
nuclear properties have been measured for $r$-process nuclei.
Several astrophysical scenarios try to explain this third abundance peak.
A detailed knowledge of nuclear lifetimes and binding energies in the
region of the N$=$126 waiting point will narrow down the possible 
astrophysical sites. If, e.g., no shell quenching could be found in this
mass range, the large dip existing for this case in front of the third 
abundance peak would have to be filled up by other processes like 
neutrino wind interactions. 
For cold decompressed neutron-rich matter, e.g., from
neutron-star mergers, we find an equilibrium between $(n,\gamma)$ and
$\beta$ decay. For the rather hot supernova-explosion scenario we 
find an equilibrium between $(n,\gamma)$ and $(\gamma,n)$ reactions. 
Considering the still rather large difficulties to identify 
convincing astrophysical sites for the third peak of the $r$-process with 
sufficiently occurrence rates, measurements of the nuclear properties 
around the N$=$126 waiting point will represent an important step forward
in solving the difficult and yet confusing site selection of the 
third abundance peak of the $r$-process.

The key bottleneck nuclei of the N$=$126  waiting point around Z$\approx 70$
are about 15 neutrons away from presently known nuclei (see Fig.~\ref{fig1}),
with a typical drop of the production cross section for classical radioactive 
beam production schemes of about a factor of 10-20 for each additional neutron 
towards more neutron-rich isotopes. 
Thus presently nothing is known about these nuclei and even next-generation 
large-scale 'conventional' radio\-active beam facilities like FAIR \cite{FAIR06},
SPIRAL II \cite{GANIL01} or FRIB~\cite{frib} will not be able to grant experimental
access to the most important isotopes on the $r$-process path. 
The third peak in the abundance curve of 
$r$-process nuclei is due to the $N=126$ waiting point as visible in Fig.~\ref{fig1}.
These nuclei are expected to have rather long halflives of a few 100~ms. 
This waiting point represents the bottleneck for the nucleosynthesis of heavy 
elements up to 
 the actinides. From the view point of astrophysics, it is the last region, 
where the $r$-process path gets close to the valley of stability and thus 
can be studied with the new isotopic production scheme discussed below. 
While the waiting point nuclei at $N=50$ and $N=82$ have been studied rather 
extensively~\cite{baruah08,dworschak08,dillmann03,blaum06}, nothing is 
known experimentally about the nuclear properties of waiting point nuclei 
at the $N=126$ magic number. Nuclear properties to be studied here are nuclear 
masses, lifetimes, $\beta$-delayed neutron emission probabilities 
$P_{\nu,n}$ and the underlying nuclear structure.

For the overall description of the $r$-process, the nuclear masses are typically
taken from mass models like the macroscopic-microscopic Finite Range Droplet 
Model (FRDM). 
Alternatively, models more closely related to first principles like
 Hartree-Fock-Bogoliubov calculations are used \cite{arnould07,kratz07}. 
Typically, somewhat less shell quenching is assumed for the heavier N$=$126 region 
compared to the N$=$82 region of the $r$-process.

\begin{figure*}[t!]
   \centerline{\includegraphics[width=.7\textwidth]{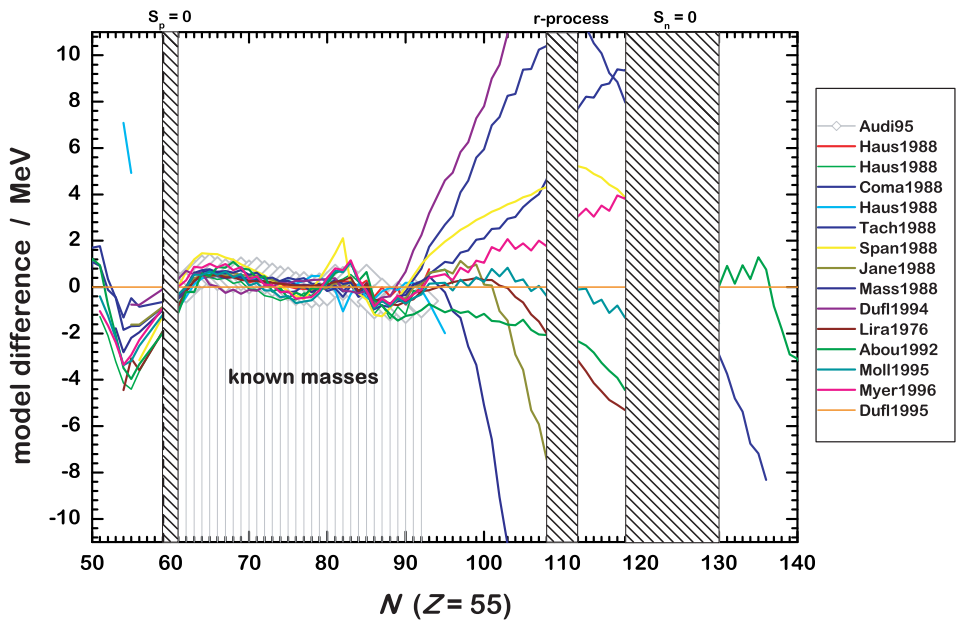}}
   \caption{Differences between nuclear mass predictions from various theoretical mass
            models for Cs isotopes (Z$=$55) compared to measured masses taken
            from AME95~\cite{audi95} as a function of the neutron number $N$
            (for $Z=$55). The figure
            is taken with permission from Ref.~\cite{blaum06}.}
   \label{fig3}
\end{figure*}

Fig.~\ref{fig3} displays the difference between nuclear masses for
the isotopic chain $Z=$55 (Cs) as calculated by various mass models
and measured masses (data taken from AME95 \cite{audi95}).
The plot demonstrates the good agreement between measured
and predicted masses in the mass range where experimental data are available,
while drastic deviations occur outside these regions especially for the 
$r$-process region~\cite{blaum06}. 
For the Cs isotopes with Z$=$55 as shown in Fig.~\ref{fig3} the mass measurements 
reach close to the $r$-process path. However, for the heavier elements 
relevant for the $r$-process waiting point N$=$126 around Z $\approx 70$, the known 
isotopes are about 15 neutrons away from the magic neutron number $N=$126 and 
Fig.~\ref{fig3} impressively illustrates the extremely large uncertainties 
presently expected for any theoretical prediction of nuclear masses, Q values, or
 $\beta$ halflives which may partly be attributed to effects of nuclear deformation.
This clearly points to the importance of direct measurements in this mass region, 
especially targeting nuclear masses.

Moreover, presently there exist difficulties to describe consistently
the third abundance peak of the $r$-process, the $^{232}$Th
and $^{238}$U cosmochronometers and the potential 'fission cycle' beyond 
$A>260$~\cite{panov05}. 

From the viewpoint of nuclear structure theory, semi-magic, heavy
nuclei not too far away from stability are spherical and thus
can be treated in general more successfully than heavy deformed nuclei.
Shell model calculations with open proton- and neutron shells require
extremely large dimensions of the configuration space.
With realistic density functional theories, some extremely time
consuming deformed RPA calculations with Skyrme or Gogny forces have
been performed~\cite{ring09}, but still spherical nuclei can be controlled
much better.
Also the subsystem of neutrons or protons allows to study systems with large
isospin.
These density functional calculations have to be fitted to
experimental data of heavy nuclei, in order to enable reliable predictions
for other nuclei~\cite{ring09}.
If we improve our experimental understanding of this final bottleneck to
the actinides at N$=$126, many new visions open up: (i) For many mass formulas
(e.g. \cite{moeller97}), there is a branch of the $r$-process
leading to extremely long-lived superheavy elements beyond Z$=$110 with
lifetimes of about $10^9$ years. If these predictions could be made
more accurate, a search for these superheavy elements in nature
would become more promising.
(ii) At present the prediction for the formation of uranium and thorium
elements in the $r$-process is rather difficult, because there are no
nearby magic numbers and those nuclei are formed during a fast
passage of the nuclidic area between shells. Such predictions could be
improved, if the bottleneck of actinide formation would be more reliably
known. (iii) Also the question could be clarified if fission
fragments are recycled in many $r$-process loops or if only a small
fraction is reprocessed. \\
This description of our present understanding of
the $r$-process underlines the importance of the present project for
nuclear physics and, particularly, for astrophysics.

\section{The Fission-Fusion Reaction Process}
  \label{sec:fissfus}
 
In the following section the various ingredients enabling the new 
fission-fusion reaction scenario are outlined. First the 
'Radiation Pressure Acceleration' method of laser ion acceleration
is described with special emphasis on the 'hole-boring' mode, which allows 
to generate ultra-dense ion beams.
Consequently, collective effects are expected for the interaction of
these ion beams with solid targets, leading to a significant
reduction of the conventional electronic stopping power. Finally,
the fission-fusion reaction process based on these ultra-dense 
laser-accelerated ion beams is described and an order-of-magni\-tude
estimate for the achievable fusion yield will be presented.

\subsection{Laser Ion Acceleration}
  \label{subsec:rpa}

Laser-accelerated energetic ion beams have been
produced during the last few years from $\mu$m thick metallic foils
when irradiated by ultra-intense, short laser pulses
\cite{hegelich06,schwoerer06,teravetisyan2006}. In these experiments the
high-energy electrons produced at the front of the target penetrated the
target being opaque to the laser. At the rear side the electrons generate
an electrostatic field, which ionizes and accelerates ions from the rear
side. This acceleration mechanism was called 'Target Normal Sheath
Acceleration (TNSA)'. It was explored in many experiments at various high-intensity
laser laboratories~\cite{snavely00,clarke00,maksimchuck00,hatchett00}.
A recent review~\cite{robson07} shows that the ion energy scales proportional to
the square root of the laser intensity. Typical conversion efficiencies
from laser energy to ion energy amount to less than 1$\%$.

In the proposal of a new nuclear reaction scenario introduced in this work,
we envisage to exploit instead the new Radiation Pressure
Acceleration (RPA) mechanism for ion acceleration. It was first proposed
theoretically~\cite{robinson09,macchi05,klimo08,robinson08,rykovanov08}. 
Special emphasis has been given to RPA with circularly polarized laser
pulses as this suppresses fast electron generation and leads to the interaction
dominated by the radiation pressure~\cite{robinson09,macchi05}.
It has been shown that RPA operates in two modes. In the first one, 
called 'hole-boring', the laser pulses interact with targets thick enough
to allow to drive target material ahead of it as a piston, but without
interacting with the target rear surface~\cite{robinson09}. 

An alternative 
scenario, called 'light-sail' (LS) mode of RPA, occurs if the target is
sufficiently thin for the laser pulse to punch through the target and accelerate
part of the plasma as a single object~\cite{klimo08,robinson08}.
Typically the 'hole-boring' mode leads to lower velocities of 
the accelerated ions, as envisaged for the present proposal.\\
The first experimental observation of RPA in the 'hole-boring' regime was 
achieved only recently in experiments
led by the Munich group~ \cite{henig09,tajima09}.

The RPA laser ion acceleration mechanism in general provides a much larger efficiency
for the conversion from laser energy to ion energy and allows for a generation
of much larger ion energies in comparison to TNSA. Moreover, for circularly
polarized laser light RPA holds promise of quasi-monoenergetic ion beams.
Due to the circular polarization, electron heating is strongly suppressed.
The electrons are compressed to a dense electron sheet in front of the laser
pulse, which then via the Coulomb field accelerates the ions.
This mechanism  requires much thinner targets and ultra-high contrast laser
pulses to avoid the pre-heating and expansion of the target before the
interaction with the main laser pulse.

The RPA mechanism allows to produce ion bunches with solid-state density
(10$^{22}$ - 10$^{23}$/cm$^3$), which thus are $\approx 10^{14}$ times more dense
than ion bunches from classical accelerators.
Correspondingly, the areal densities of these bunches are $\approx 10^7$
times larger.
It is important to note that these ion bunches are accelerated as neutral
ensembles together with the accompanying electrons and thus do not Coulomb
explode.

For an estimate of the required laser intensities, focal spot area
and target thickness, the 1-D RPA
model as outlined in~\cite{robinson09} is sufficient. It holds true for
the relativistic 'hole-boring' regime of RPA. For the achievable ion energy $E_i$
it yields the expression (circular polarized light)

\begin{equation}
  E_i = E_u \cdot A= 2m_i c^2 \Xi / \left( 1 + 2 \sqrt{\Xi} \right)  ,
  \label{ion-energy}
\end{equation} 

where $E_u$ is the energy per nucleon, A is the atomic mass number, 
$m_i$ is the ion mass, $c$ is the vacuum speed of light, and $\Xi$ is 
the dimensionless pistoning parameter given by

\begin{equation}
    \Xi = I_L / (m_i n_i c^3) .  
 \label{piston}
\end{equation}

$I_L$ denotes the laser intensity and $n_i$ the ion density. In the 
non-relativistic limit $\Xi << 1$, Eq.~(\ref{ion-energy}) reduces 
to $E_i = 2 m_i c^2 \Xi$, which together with Eq.~(\ref{piston}) is
equivalent to Macchi's Eq.~(1) in~\cite{macchi05}. 
The conversion efficiency of laser energy to ion 
energy, $\chi$, follows from~\cite{robinson09}

\begin{equation}
  \chi = 2 \sqrt{\Xi} / \left( 1 + 2 \sqrt{\Xi} \right)  .
  \label{chi}
\end{equation}

The total number of ions, $N_i$, that can be accelerated results from the 
energy balance

\begin{equation}
    N_i E_i = \chi W_L  , 
 \label{energy-balance}
\end{equation}

where $W_L$ denotes the energy of the laser pulse.

The target arrangement we want to use is depicted in Fig.~\ref{fig4}. 
It actually consists of two targets termed production target and reaction 
target. The first is composed of two spatially separated foils, one made from 
thorium and the other from deuterated polyethylene, CD$_2$. They serve for the 
generation of a thorium ion beam and a beam containing carbon ions and 
deuterons. The reaction target has a sandwich structure. The first layer is 
made from CH$_2$ and causes fission of the accelerated thorium nuclei. The 
second layer is a pure thorium film. The accelerated carbon ions and deuterons 
lead to fission of these thorium nuclei. Fusion of the fragments created in 
both layers generates neutron-rich nuclei in a mass range towards the waiting 
point N$=$126. This reaction scheme works best when the thorium and carbon ions 
and the deuterons have each the energy of 7 MeV per nucleon (for details see 
Sect.~\ref{subsec:stopping} and the following sections). \\    
Accelerating $^{232}$Th ions whose density $\rho_{\rm Th}= m_{\rm Th}n_{\rm Th}$ 
amounts to 11.7 g/cm$^3$ to $E_u$=7 MeV per nucleon with laser light of 0.8 $\mu$m 
wavelength needs, according to equations (1) and (2), an intensity 
of 1.2$\cdot 10^{23}$ W/cm$^2$. The dimensionless vector potential, $a_L$, follows 
from

\begin{equation}
    a_L=\sqrt{\frac{f \cdot I_L[Wcm^{-2}]\cdot \lambda_L^2[\mu m^2]}
                   {1.37\cdot 10^{18}}} 
 \label{a_L}
\end{equation}

with $f=1$ for linear and $f= 1/2$ for circular polarized light, respectively.
Eq.~(\ref{a_L}) gives the value of 167 for 1.2$\cdot 10^{23}$ W/cm$^2$ and $\lambda_L=0.8\mu m$,
at circularly polarized light. The conversion efficiency, $\chi$, reaches 
11$\%$ ($\Xi = 3.8 \cdot 10^{-3}$). Intensities of this level will be 
achievable with the APOLLON facility, which is under development at the 
ENSTA/Ecole Polytechnique in Palaiseau within the ILE project~\cite{ILE}
and will form the backbone of the ELI Nuclear Physics project. The APOLLON 
single-beam pulses will provide $W_L=$~150 J in $t_L=$~32 fs, 
corresponding to 4.7 PW.
The sum of two of these beams is assumed to be available for the present estimate. 
Because of $W_L = I_L \cdot A_F \cdot t_L$ these values fix 
the focal spot area on the thorium production target, $A_F$, to 
7.1 $\mu$m$^2$ (3 $\mu$m diameter) and, from Eq.~(\ref{energy-balance}) the 
number of accelerated 
thorium ions, $N_i$, to 1.2$\cdot 10^{11}$. The thickness of the thorium 
foil, $d_{\rm Th}$, follows from $N_i= A_F d_{\rm Th} n_{\rm Th}$ and amounts 
to 560 nm ($n_{\rm Th} = 3 \cdot 10^{22}/cm^3$). \

The data for the CD$_2$ case is obtained similarly. As shown in~\cite{robinson09a}, 
the carbon ions and deuterons will experience the same energy per nucleon. 
The pistoning parameter and the conversion efficiency have hence the same values 
as before, $\Xi = 3.8 \cdot 10^{-3}$ and $\chi =$ 0.11. Eq.~(\ref{piston}) then 
yields $1.0 \cdot 10^{22}$ W/cm$^2$ ($a_L =$48) for the focal intensity, $I_L$, 
whereby for the polyethylene density, $\rho_{\rm PE} = m_cn_c + m_d n_d$, the 
value of 1 g/cm$^3$ is taken. Assuming here again a focal spot diameter of 
of 3 $\mu$m ($A_F =$~7.1 $\mu$m$^2$), the required laser 
energy, $W_L = I_L A_F t_L$, results in 23 J. 
The number of accelerated carbon ions and deuterons amounts 
to $1.4 \cdot 10^{11}$ and $2.8 \cdot 10^{11}$, respectively. The thickness of 
the polyethylene foil, $d_{\rm PE}$, is 520~nm.
  
Phase-stable acceleration~\cite{yan08,tripathi09} would yield 
mono\-chroma\-tic ion energy 
spectra. Whether this can be really achieved, in particular when several 
ion components with different charge-to-mass ratios are present, is 
hardly predictable on the basis of current experimental and theoretical 
knowledge on ion acceleration. So for a safe evaluation of the fusion 
process of the thorium fragments, the ion spectra are assumed to be 
broad (see Sect.~\ref{subsec:fissfus}).   \\
Predictions related to the important question of the beam stability 
based on 2-D or 3-D simulations show that plane foils heavily expand 
and break up due to the Ray\-leigh-Taylor instability~\cite{yu10}. Promising 
counter-measures include targets adequately modulated in density and 
shape ~\cite{chen09,yan08}.


\subsection{Stopping Power for Dense Ion Bunches in a Solid Target}
 \label{subsec:stopping}

In nuclear physics the Bethe-Bloch formula~\cite{segre77}
is used to calculate the atomic stopping of energetic individual ions

\begin{equation}
    - \frac{dE}{dx} = 4\pi n_e \frac{Z_{eff}^2e^4}{m_ev^2}
            \left[ ln \left\{ \frac{2m_e v^2} {I_p(1-(v/c)^2)} \right\} 
              - \left( \frac{v}{c} \right)^2 \right],
\end{equation}

where $I_p$ denotes the ionization potential, $n_e$ the electron density,
$m_e$ the mass of the electron, while $v$ is the ion velocity and $Z_{eff}$ is
the effective charge of the ions.

For laser-accelerated ions the ion bunch densities reach solid-state density, 
which is about 14 orders of magnitude larger
compared to beams from classical accelerators.
In such a scenario collective effects become important.
According to Ref.~\cite{ichimaru73}, the Bethe-Bloch equation can be decomposed
into a first part describing binary collisions and a second
term describing long-range collective contributions according to

\begin{equation}
     {-dE/dx}=4\pi n_e \frac{Z_{eff}^2e^4}{m_ev^2}[ ln (m_ev^2/e^2k_D)
          + ln(k_Dv/\omega_p)].
\end{equation}

Here $k_D$ is the Debye wave number and $\omega_p$ is the plasma frequency
of the electrons. In Ref.~\cite{wu09} the mechanism of collective
deceleration of a dense particle bunch in a thin plasma is discussed, where
the particle bunch fits into a half of one plasma oscillation and is
decelerated $10^5-10^6$ times stronger than predicted by the
classical Bethe-Bloch equation \cite{segre77} due to a strong collective
wakefield.
Now we discuss the opposite effect with a strongly reduced atomic
stopping power that occurs when sending the energetic, solid-state density
ion bunch into a solid carbon or thorium target.
For this target the plasma wavelength ($\lambda_p\approx$5 nm, driven
by the ion bunch with a phase velocity corresponding to the thorium ion 
velocity) is much
smaller than the ion bunch length ($\approx$ ~560~nm) and collective
acceleration and deceleration effects cancel. Only the binary collisions
remain and contribute to the stopping power. Hence, we may consider the
dense ion bunch as consisting of about 1750 atomic layers with 
a distance between the Th ions of about 3.2{\AA} as obtained from the bulk 
density of metallic thorium (11.7 g/cm$^3$).
In this case the first layers of the ion bunch will attract the electrons
from the target and like a snow plough will take up the decelerating electron
momenta. Hence the predominant part of the ion bunch is
screened from electrons and we expect a drastic reduction of the stopping power.
The electron density $n_e$ will be strongly reduced in the
channel defined by the laser-accelerated ions, because many electrons are expelled
by the ion bunch and the laser pulse.
This effect requires detailed experimental investigations planned for the near
future, aiming at verifying the perspective to use a significantly thicker
reaction target. The classical ion range for
e.g. 7 MeV/u thorium ions in carbon is 15 mg/cm$^2$,
corresponding to a range of 66 $\mu$m, while this range amounts to only
40~$\mu$m in a thorium target. However,
if we aim at limiting the usable effective range to a thorium target thickness
where the remaining projectile energy is still sufficient to induce fission,
using the accelerated thorium ions directly to induce fission in the Th target
would result in a usable target range of less than 10~$\mu$m without 
invoking collective effects. However, the use of proton induced fission
leads to a usable target thickness of about 50 $\mu$m.\\
The expected reduced atomic stopping power will be supported by the strong laser
heating of the electrons. A reduction of the atomic stopping is essential to avoid
a strong slowing down of the ions below the Coulomb barrier energies,
where nuclear reactions are no longer possible. However, even without
this reduced stopping power the basic properties of the novel reaction
mechanism could still be studied, but with significantly reduced yields.\\
Taking collective effects into account by assuming a range enhancement by
a factor of 100, we expect a usable thickness of several mm for a thorium
target.\\
An optimized ion acceleration scheme will depend on measured stopping 
powers of the dense bunches in targets of different materials and thicknesses,
including the ion beam energy as a further parameter to be optimized in 
preparatory studies.

\subsection{The Fission-Fusion Process}
  \label{subsec:fissfus}

The basic concept of the fission-fusion reaction scenario draws on the
ultra-high density of laser-accelerated ion bunches. Choosing fissile
isotopes as target material for a first target foil accelerated by an
intense laser pulse will enable the interaction of a dense beam of
fission fragments with a second target foil also consisting of
fissile isotopes. So finally in a second step of the reaction process,
fusion between (neutron-rich) beam-like and target-like fission products 
will become possible, generating extremely neutron-rich ion species.

For our discussion we choose $^{232}$Th (the only component of chemically 
pure Th) as fissile target material, primarily because of its long halflife 
of 1.4$\cdot 10^{10}$ years, which avoids extensive radioprotection 
precautions during handling and operation. 
Moreover, metallic thorium targets are rather stable in a typical 
laser vacuum of $10^{-6}$ mbar, whereas e.g. metallic $^{238}$U targets 
would quickly oxydize. \\
Nevertheless, in a later stage it may become advantageous
to use also heavier actinide species in order to allow for the production
of even more exotic fusion products.

In general, the fission process of the two heavy Th nuclei from beam and 
target will be preceded by the deep inelastic transfer of neutrons between the 
inducing and the fissioning nuclei. Here the magic neutron number in the 
superdeformed fissile nucleus with N$=$146~\cite{thirolf02,metag75}
may drive the process towards more neutron-rich fissioning nuclei,
because the second potential minimum acts like a doorway state towards fission.
Since in the subsequent fission process the heavy fission fragments keep their
$A$ and $N$ values~\cite{vandenbosch73}, these additional neutrons will
show up in the light fission fragments and assist to reach more
neutron-rich nuclei. This process will be of particular importance in the
reaction scenario discussed in Sect.~\ref{subsec:collective} for the case
of collectively reduced stopping in the reaction target.

\begin{figure*}[t!]
   \centerline{\includegraphics[angle=-90,width=.7\textwidth]{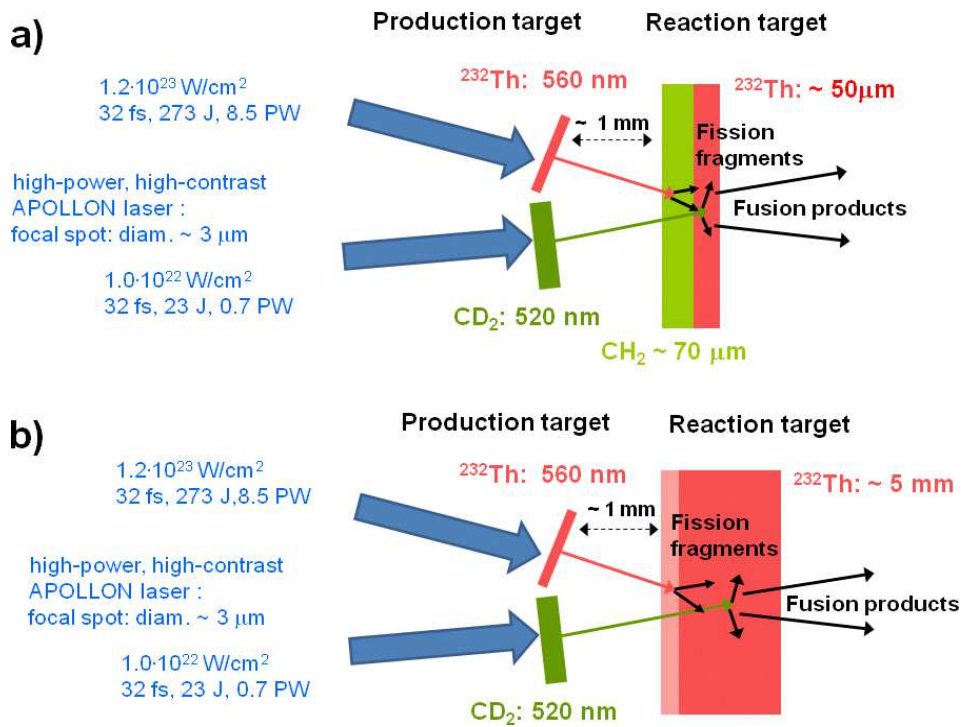}}
   \caption{Sketch of the target arrangement envisaged for the fission-fusion
            reaction process based on laser ion acceleration. Part a) illustrates
            the situation in case no collective effects on the electronic stopping
            are taken into account. In this case the thickness of the CH$_2$ layer 
            as well as the second thorium reaction target have to be 
            limited to 70 $\mu$m and 50 $\mu$m, respectively, in order to enable 
            fission of beam 
            and target nuclei. This will allow for fusion between their light 
            fragment as 
            well as enable the fusion products to leave the second thorium 
            reaction target. Part b) depicts an alternative scenario, where
            we consider also collective effects in the reaction target
            induced by the ultra-dense ion bunches. Here the first part of
            the thorium reaction target is used to decelerate the fission fragments
            from about 7 MeV/u to about 3 MeV/u, suitable for efficient fusion
            of neutron rich species.
            Due to the reduced electronic stopping a larger target thickness and
            thus increased fission and fusion yields can be expected. Further 
            details are discussed in the text.}
   \label{fig4}

\end{figure*}

Fig.~\ref{fig4} shows a sketch of the proposed fission-fusion reaction scenario
for two different situations, a) for the case of normal electronic stopping as
described by the Bethe-Bloch equation and b) for the case of reduced stopping
due to collective effects in the target induced by the ultra-dense ion beam
discussed earlier. The latter scenario will be discussed later. \\

As mentioned before, the accelerated thorium ions are fissioned in the 
CH$_2$ layer of the reaction target, whereas the carbon ions and deuterons
generate thorium fragments in the thick thorium layer of the reaction target.
This scenario is more efficient than the one where fission would be induced
by the thorium ions only.

For practical reasons we propose to place the reaction
target about 1~mm behind the production target, as indicated in Fig.~\ref{fig4}.

\subsubsection{Induced fission with normal electronic stopping}
  \label{subsec:normstopp}

In the scenario, where the earlier discussed collective effects
in the target are not taken into account (marked with 'a)' in Fig.~\ref{fig4}), 
the thorium layer of the reaction target would have a thickness of about 50~$\mu$m.

Using a distance of 2.8 $\AA$ between atoms in solid layers of CH$_2$, 
the accelerated light ion bunch (1.4$\cdot 10^{11}$ ions) corresponds 
to 1860 atomic layers in case 
of a 520~nm thick CD$_2$ target. In order to allow for an optimized fission of 
the accelerated Th beam, the thicker Th layer of the reaction target, 
which is positioned behind the production target, is covered by 
about 70~$\mu$m of polyethylene. 
This layer serves a twofold purpose: Primarily it is used to induce fission of the 
impinging Th ion beam, generating the beam-like fission fragments. Here polyethylene
is advantageous compared to a pure carbon layer because of the increased number
of atoms able to induce fission on the impinging Th ions.
In addition, the thickness of this CH$_2$ layer has been chosen such 
that the produced fission fragments will be decelerated to a kinetic energy 
which is suitable for optimized 
fusion with the target-like fission fragments generated by the light accelerated 
ions in the Th layer of the reaction target, minimizing the amount of 
evaporated neutrons. 
After each laser shot, a new double-target has to be rotated into position. \\

In order to estimate the fission cross sections both of beam and target nuclei, we
apply geometrical considerations based on the involved nuclear radii, which 
can be expressed for mass number $A$ in the usual way as

\begin{equation}
    R=1.2 \cdot (A)^{1/3} fm.
\end{equation}

Neglecting the influence of surface diffuseness effects, the resulting fission cross 
section of the $^{232}$Th beam in the CH$_2$ layer of the reaction target 
amounts to $\sigma_{\rm fis}= \pi (R_1 + R_2)^2 = 350~fm^2 = 3.5\cdot 10^{-28}~m^2$
(3.5~b). 
Correspondingly, the deuteron-induced fission in the Th reaction target occurs with a 
cross section of about 247~fm$^2 = 2.47\cdot 10^{-28}$m$^2$ (2.47~b). If we use 
the atomic distance of 3.2~{\AA}~for thorium, we conclude a fission 
probability of about 4.1$\cdot 10^{-9}$ per atomic layer.\\
In order to estimate the required thickness of the CH$_2$ front layer
of the reaction target, we have to take into account the range of the  
7 MeV/u $^{232}$Th ions, which is about 120~$\mu$m. However, already after 70~$\mu$m the
kinetic energy of the Th ions has dropped to 3 MeV/u, which is about the energy
required for the resulting fission fragments during the subsequent fusion step.
Therefore, we estimate the thickness of the polyethylene layer to about 70~$\mu$m, 
which corresponds to $\sim$ 2.5$\cdot 10^5$ atomic layers.
Together with the above estimated fission probability per atomic layer and taking into
account that from CH$_2$ three atoms will contribute to the fission process of the
impinging Th beam, this results in a fission probability for the Th ion beam 
of about 3.1$\cdot 10^{-3}$ in 
the 70~$\mu$m CH$_2$ layer, thus generating about 3.7$\cdot 10^8$ 
beam-like fission fragments per laser pulse. \\ 
The 99.7$\%$ of Th beam ions passing through the CH$_2$ layer will enter 
the Th layer of the reaction target
with about 2.4 MeV/u, corresponding to a range of 21~$\mu$m. In the first atomic 
layers a fraction of them will undergo fission before being slowed down too much, 
however, the resulting fragment energies will not be suitable for the fusion step.
A quantitative assessment of this component would require detailed simulations and 
will be finally addressed by experimental studies.

In general, the fission process proceeds asymmetric~\cite{vandenbosch73}. The heavy 
fission fragment for $^{232}$Th is centred at A=139.5 (approximately at Z=54.5 
and N=84) close to the magic numbers Z=50 and N=82. Accordingly, the light 
fission fragment mass is adjusted to the mass of the fixed heavy fission fragment,
thus resulting for $^{232}$Th in $A_L$=91 with $Z_L\approx 37.5$. 
During the fission process of $^{232}$Th for low excitation energies, 
typically 1.6 neutrons are emitted. However, for the discussion presented here we 
neglect this loss of neutrons, because 4 or 5 neutrons may be transfered to the 
fissioning nucleus in the preceding transfer step (particularly efficient and
thus important in case of the Th-induced fission discussed in the following section). 
The width (FWHM) of the light 
fission fragment peak is typically $\Delta A_L=14$ mass units, the $1/10$ maximum 
width about 22 mass units~\cite{vandenbosch73}.\\

So far we have considered the fission process of beam-like Th nuclei in the 
CH$_2$ layer of the reaction target. Similar arguments can be invoked for the
deuteron- (and carbon) induced generation of (target-like) fission products in 
the subsequent thicker thorium layer of the reaction target,
where deuteron- and carbon-induced fission will occur in the $^{232}$Th layer 
of the reaction target. 
Since we can consider the 2.8$\cdot 10^{11}$ 
laser-accelerated deuterons (plus 1.4$\cdot 10^{11}$ carbon ions) impinging 
on the second target per laser pulse as 1860 consecutive 
atomic layers, we conclude a corresponding fission probability in the Th layer 
of the reaction target of about 2.3$\cdot 10^{-5}$, corresponding 
to 3.2$\cdot 10^6$ target-like fission fragments per laser pulse. 
A thickness of the 
thorium layer of the reaction target of about 50~$\mu$m could be exploited, where the 
kinetic proton energy would be above the Coulomb barrier to induce fission 
over the full target depth. \\

An essential effect to be taken into account is the widening of the fission 
fragment beam, because a kinetic energy of about 1 MeV/u is added to the fission 
fragments in arbitrary directions. However, the angular distribution of fission 
fragments from proton (or heavy ion) induced fission follows a 1/sin($\Theta$) 
distribution~\cite{vandenbosch73} (with $\Theta$ denoting the fragment angle 
with respect to the direction of the incoming beam inducing the fission process) 
and thus fragments are predominantly 
emitted in beam direction. Consequently, a fraction of a few percent will stay 
within the conical volume defined by the spot diameter of the laser 
focus on the production target.\\
Due to the additional kinetic energy of about 1 MeV/u of the fission fragments 
in the thick reaction target also the target-like fragment volume will expand. 
Here the very short bunch length of the fragment beam becomes important. 
The beam velocity is about 10$\%$ of the velocity of light and during the short
fly-by time of the ions of only 1 fs the fission fragments of the target
can only move a distance of 1 $\mu$m, which is still small compared to the beam
diameter of 3~$\mu$m. Thus this enlargement of the target area is negligible.

In a second step of the fission-fusion scenario, we consider the fusion between 
the light fission fragments of beam and target to a compound nucleus
with a central value of $A\approx 182$ and $Z\approx 75$.\\
Again we employ geometrical arguments for an order-of-magnitude estimate
of the corresponding fusion cross section. For a typical light fission
fragment with $A=90$, the nuclear radius can be estimated as 5.4~fm.
Considering a thickness of 50~$\mu$m for the Th layer of the reaction target 
that will be
converted to fission fragments, equivalent to 1.6$\cdot 10^5$ atomic layers, 
this results in a fusion probability of about 1.8$\cdot 10^{-4}$.

With this estimate for the fusion cross section, we can finally derive an 
order-of-magnitude for the final yield of fusion products generated via the
presented fission-fusion process of about 1-2 fusion products per
laser shot. This estimate does not yet take into account any collective
effects in the target that might result in much higher fission rates and
accordingly increased fusion yields as discussed in the following section. 

Besides the fusion of two light fission fragments other reactions may happen.
The fusion of a light fission fragment and a heavy fission fragment would lead
back to the original Th nuclei, with large fission probabilities, thus we can 
neglect these fusion cross sections.
The fusion of two heavy fission fragments would lead to nuclei
with A$\approx 278$, again nuclei with very high fission probability.
Hence we have also neglected these rare fusion cross sections, although 
they may be of interest on their own. Thus we concentrate here only on the 
fusion of two light fission fragments. Besides studying nuclei close to the 
waiting point of the $r$-process with the magic neutron number N$=$126, 
we may investigate also neutron-rich isotopes with the magic proton number 
Z=82, which are of large interest in nuclear structure studies.

Very neutron-rich nuclei still have comparably small production cross sections, 
because weakly bound neutrons ($B_N \ge$ 3 MeV) will be evaporated easily.
The optimum range of beam energies for fusion reactions resulting in 
neutron-rich fusion products amounts to about 2.8 MeV/u according to PACE4~\cite{pace} 
calculations. So, e.g., the fusion of two neutron-rich $^{98}_{35}$Br fission products 
with a kinetic energy of the beam-like fragment of 275 MeV leads with an excitation
energy of about 60 MeV to a fusion cross section of 13 mb for 
$^{189}_{70}$Yb$_{119}$, which is already 8 neutrons away from the last presently
known Yb isotope.\\
One should note that the well-known hindrance of fusion for nearly symmetric systems
(break-down of fusion) only sets in for projectile and target masses 
heavier than 100~amu~\cite{quint93,morawek91}. Thus for the fusion of light 
fission fragments, we expect an unhindered fusion evaporation process. 

\begin{figure*}[t!]
   \centerline{\includegraphics[angle=-90,width=.7\textwidth]{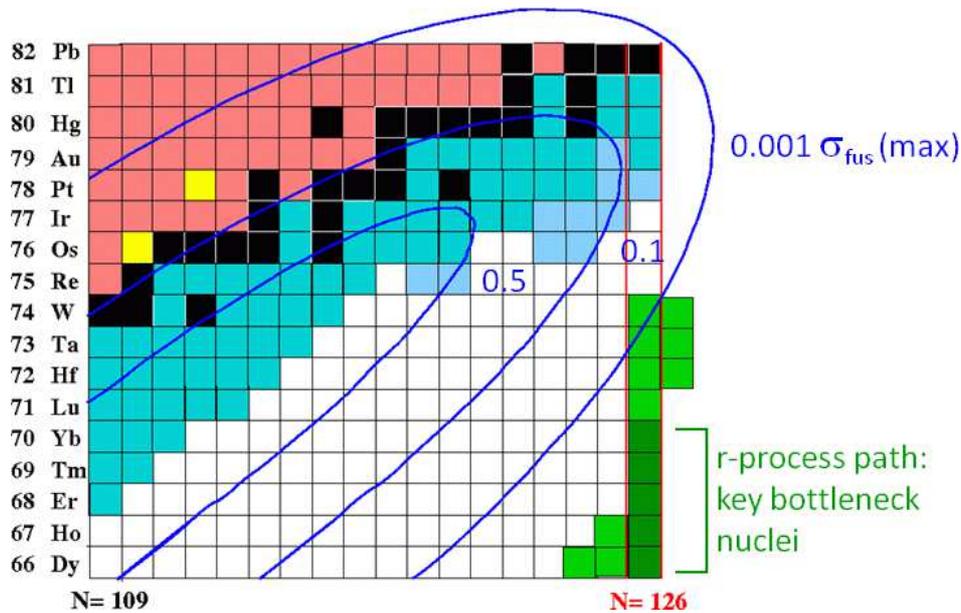}}
   \caption{Chart of nuclides around the N$=$126 waiting point of the
            $r$-process path. The blue ellipses denote the expected range
            of isotopes accessible via the novel fission-fusion process.
            The indicated lines represent 0.5, 0.1 and 0.001 of the maximum
            fusion cross section after neutron evaporation. In green the
            N$=$126 nuclides relevant for the $r$-process are marked, with
            the dark green colour indicating the key bottleneck nuclei
            for the astrophysical $r$-process.}
   \label{fig5}
\end{figure*}

In Fig.~\ref{fig1} the range of reachable fusion products from the 
fission-fusion process is indicated by the blue ellipses overlayed to the
chart of nuclides.
The proton to neutron ratio, which is approximately conserved during fission
(indicated by the straight blue line connecting $^1$H with $^{238}$U) 
determines the slope of the inner blue elliptical contour in Fig.~\ref{fig1}.
Their eccentricity reflects the region of nuclei reachable within a range 
of 50$\%$
of the maximum fusion cross section, based on the large fluctuations
of proton and neutron numbers of the participating fission fragments.
So far the dimensions of the contour lines 
drawn in Fig.~\ref{fig1} and Fig.~\ref{fig5} have been estimated from usual 
fission distributions. Since the mass distributions of both light fission 
fragments exhibit a certain width (FWHM), the width of the resulting 
distribution after fusion will be about a factor of $\sqrt{2}$ larger. 
Those distributions will steepen when reaching further out to their tails. 
The other two elliptical contour lines correspond to the regions of fusion 
products expected to be reachable with 10$\%$ and $10^{-3}$ of the maximum 
cross sections, respectively.\\
Fig.~\ref{fig5} displays a closer view into the region of nuclides around
the N$=$126 waiting point of the $r$-process, where nuclei on the $r$-process 
path are indicated by the green colour, with dark green highlighting the 
key bottleneck $r$-process isotopes~\cite{risac} at N$=$126 between Z$=$66 (Dy) and 
Z$=$70 (Yb). One should note that, e.g., for Yb the presently last known
isotope is 15 neutrons away from the $r$-process path at N$=$126.
The isotopes in light blue mark those nuclides, where recently
$\beta$-halflives could be measured following projectile fragmentation
and in-flight separation 
at GSI~\cite{kurtu08}. Again the elliptical contour lines indicate the range 
of nuclei accessible with our new fission-fusion scenario on a level of
50$\%$, 10$\%$ and 10$^{-3}$ of the maximum fusion cross section between
two neutron-rich light fission fragments in the energy range of about 2.8 MeV/u, 
respectively.

\subsubsection{Fission-Fusion with collectively reduced electronic stopping}
  \label{subsec:collective}

So far we estimated the expected fission and fusion yield without
refering to any collective effects in the reaction targets that may
reduce the electronic stopping as discussed earlier. Now we extend
this discussion by considering the expected reduction of the electronic stopping 
in the reaction target.
This scenario would allow to extend the thickness of the Th production target
to probably a few mm (situation 'b)' in Fig.~\ref{fig4}).\\
While so far no experimental data or quantitative assessment on the amount
of collective range enhancement is available, we assume for the discussion
within this paragraph a factor of 100 and discuss the consequences.

In contrast to the previously discussed scenario without collective effects,
we now propose to abandon the front carbon layer of the reaction target
and use only a homogeneous, thick Th target as indicated in Fig.~\ref{fig4}b).
In this case we use the first part of the target primarily as stopping medium
for the incoming energetic Th ions in order to decelerate them from initially 
about 7 MeV/u to about 3 MeV/u, which is suitable for the subsequent fusion
step with target-like fragments from proton-induced fission. Since such a 
deceleration could be reached in about 16~$\mu$m without collective effects, 
we estimate here about 0.2~mm from our 5~mm thick Th reaction target acting as
stopper while producing fission fragments too fast for efficient fusion of 
extremely neutron-rich isotopes. This part of the reaction target is marked 
by the lighter red colour in Fig.~\ref{fig4}b). However, this part amounts to 
only 4$\%$ of the reaction target and thus does not lead to a significant loss of 
usable fission yield. On the other hand, neutron transfer towards the deformed
neutron shell closure at N$=$146 preceding fission will add 4 neutrons to the
light fission fragment and thus significantly help to enhance fusion of very 
neutron-rich isotopes. 
Compared to the situation of Fig.~\ref{fig4}a), a thickness of the thorium target 
increased by a factor of 100 to about 5~mm due to the correspondingly reduced 
stopping would result in a full conversion of the Th beam into fission fragments 
(with $\sim 96\%$ in an energy range usable for the fusion step).
Thus 1.2$\cdot 10^{11}$ beam-like light fission fragments would become available for
the fusion stage of the reaction process. \\
Consequently the deuteron-induced fission yield in the reaction target would also 
rise by the same factor of 100 to a target fission 
probability of 2.3$\cdot 10^{-3}$. Here we use the assumption of a linear rise of the 
energy loss with target thickness, preserving a kinetic proton energy above the
fission barrier to induce fission over the full target depth. 

So we conclude that the expected collective stopping range enhancement
will lead to a drastic increase of the fusion yield from about 1-2
fusion products per laser pulse to a value of about $4\cdot 10^4$ exotic nuclides
per pulse. Most likely only part of this estimated yield enhancement could finally 
be realized, so it may be adequate to finally quote the average between the 
two extremes, resulting in an estimate of about 10$^3$ fusion products generated
per laser pulse. However, it is obvious that collective effects from the ultra-dense 
ion bunches would significantly improve the experimental conditions towards the 
production of extremely neutron rich fusion products.
Moreover, if we could use a layered production target instead of the presently
separated arrangement, while still achieving quasi-monoenergetic
thorium and CD$_2$ ion beams, then the primary fission fragment rates would each 
increase by a factor of 2, thus resulting in an increase of the fusion yield by a 
factor of 4.

The following table finally gives a quantitative overview of the two discussed
experimental scenarios with and without collective stopping reduction, based
on the parameters of the driver laser introduced earlier.
All numbers refer to yields expected for one laser pulse.

\begin{table}
  \caption{Compilation of relevant parameters determining the
           expected yield (per laser pulse) of the fission-fusion 
           reaction process proposed in this work.}
\begin{tabular}{l|l|l}
    \hline\noalign{\smallskip}
        & normal & reduced   \\
        & stopping & stopping  \\
     \noalign{\smallskip}\hline\noalign{\smallskip}
  production target:   &                     &          \\
  $^{232}$Th            &  560 nm          & 560 nm     \\
  CD$_2$                &  520 nm          & 520 nm  \\
  \hline
  accelerated Th ions  & 1.2 $\cdot 10^{11}$ & 1.2 $\cdot 10^{11}$   \\
  accelerated deuterons& 2.8 $\cdot 10^{11}$ & 2.8 $\cdot 10^{11}$ \\
  accelerated C ions   & 1.4 $\cdot 10^{11}$ & 1.4 $\cdot 10^{11}$ \\
  \hline
  reaction target:     &                    &                   \\
  CH$_2$               & 70~$\mu$m  &  --  \\
  $^{232}$Th           & 50~$\mu$m  & 5~mm     \\
  \hline
  beam-like light fragments   & 3.7 $\cdot 10^8$  & 1.2$\cdot 10^{11}$ \\
  target-like fission probability  & 2.3$\cdot 10^{-5}$ & 2.3$\cdot 10^{-3}$ \\
  target-like light fragments & 3.2$\cdot 10^{6}$  & 1.2$\cdot 10^{11}$ \\
  fusion probability   &   1.8 $\cdot 10^{-4}$   & 1.8 $\cdot 10^{-4}$ \\
  \hline
  fusion products      &   1.5     & 4 $\cdot 10^4$  \\
    \noalign{\smallskip}\hline
\end{tabular}
\end{table}

While it will remain a challenge to directly study the key waiting point 
isotopes on the $r$-process path, it is on the other hand intriguing that
a wide range of so far unknown isotopes will become accessible for
experimental investigation. \\

Presently the high-intensity APOLLON laser envisaged to 
be used for laser ion acceleration is designed to operate at a repetition rate of
one laser pulse per minute. However, laser technology is progressing rapidly
with large efforts presently devoted to the development of higher repetition
rates, aiming of up to 10 Hz together with an increase of the laser pulse energy
beyond 1 kJ. Moreover, since the yield of very neutron-rich 
fusion products grows strongly nonlinear with laser energy, a final use of 
several coincident APOLLON laser beams would be very advantageous. \\
Therefore, it is foreseeable that the above
given estimate for the achievable rate of neutron-rich fusion products can be 
increased within the next years significantly by several orders of magnitude.

\section{Experimental aspects}

\begin{figure*}[]
   \centerline{\includegraphics[angle=-0,width=.9\textwidth]{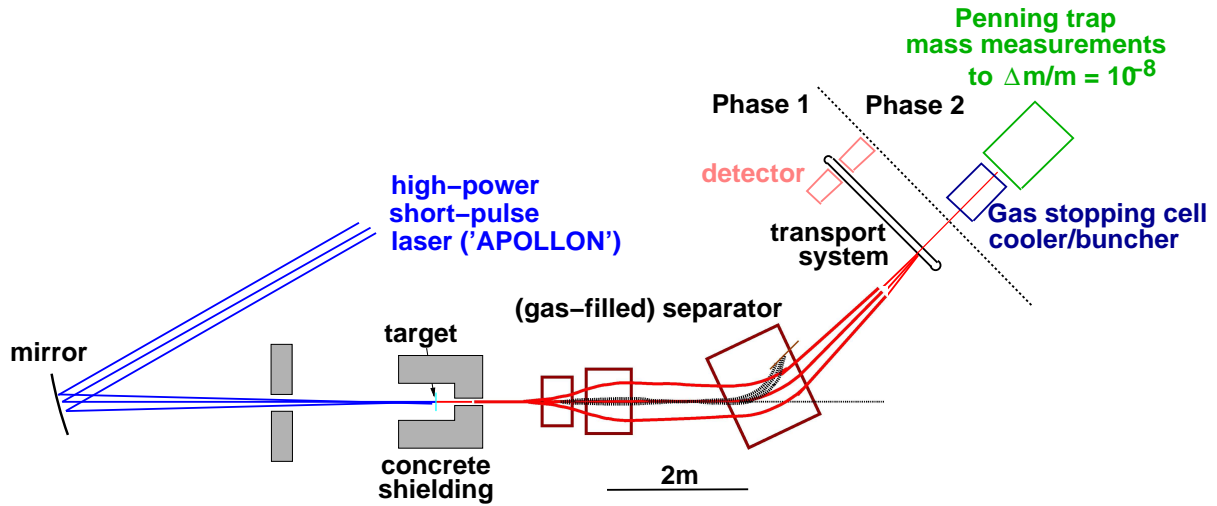}}
   \caption{Schematical view of the experimental arrangement for
            fission-fusion studies. Measurements of fusion products will be
            performed in two stages, first aiming at an identification of the
            produced isotopes via decay spectroscopy using a transport
            system (e.g. tape), while later on precision mass measurements
            using a Penning trap system are envisaged.}
   \label{fig6}
\end{figure*}

Exploring this 'terra incognita' of yet unknown isotopes towards the 
$r$-process waiting point at N$=$~126 certainly calls for a staged experimental 
approach. First studies should focus on the range and electronic stopping 
powers of dense laser-accelerated ion beams as discussed previously, 
followed by systematic optimizations of target properties in order to 
optimize the yield of fission fragments. 
Subsequently the A, Z and N distributions of the 
light thorium fission fragments should be characterized.
Moreover, it is unclear in how far the first neutron transfer preceding 
fission will additionally broaden all these distributions. 
Also the yields for the fusion products should be measured in exploratory 
experiments, where it will be crucial to optimize the kinetic energy of the 
beam-like fission products. 

Fig.~\ref{fig6} shows a schematical view of the potential experimental
setup of the presented reaction scenario. The high-intensity laser beam
is tightly focussed onto the target assembly. This area will require
heavy concrete shielding for radioprotection. The probably most essential and 
also most demanding experimental task will be the separation of the reaction
products. Fusion products with about 2-3 MeV/u will have to
be separated from faster beam-like fission fragments with about 7 MeV/u, 
or target-like fragments with about 1 MeV/u, which could 
be achieved with a velocity filter. However, the reaction products from various 
fusion channels varying in mass but not in velocity require a different 
separation scheme. Here one could use a recoil separator (as indicated in
Fig.~\ref{fig6}), where it may be advantageous to operate the separator
in gas-filled mode. Alternatively, also a coarse magnetic dipole pre-separator
followed by a gas stopping cell and an RFQ cooler/buncher could be used
to inject the ions into an electrostatic mass separator
like the 'Multi-Reflection Time-of-Flight' mass spectrometer~\cite{plass08}, 
especially when aiming for fusion products with lifetimes
significantly shorter than 100~ms. Such a spectrometer could be operated
either as an isobar separator or directly for mass measurements with
a mass accuracy of up to $10^{-7}$.

In these first studies, a tape station could be used to transport the 
reaction products to a remote, well-shielded detector system, where the
characterization of the implanted fusion products could be performed
either via $\beta$-decay studies using, e.g., LaBr$_3$ scintillation
detectors or $\gamma$ spectroscopy with high-resolution germanium detectors.
This scenario has been labelled 'Phase 1' in Fig.~\ref{fig6}.
Since most of the fusion products have typical lifetimes of $\approx 100$ ms,
they will survive the transport to a secondary target and/or detector station.

In a later stage ('Phase 2'), the fusion products may be stopped in a buffer gas 
stopping cell~\cite{neumayr07}, cooled and bunched in, e.g., a radiofrequency
quadrupole ion guide before then being  transferred to a Penning trap system
for high-accuracy mass measurements. Such a setup would be similar to the 
SHIPTRAP facility at GSI~\cite{block05} or ISOLTRAP at ISOLDE/CERN~\cite{mukh05} 
for mass measurement with an accuracy of $\Delta$m/m$\sim 10^{-8}$,
corresponding to about 10 keV/c$^2$ \cite{block10}.

\section{Conclusion}

The exploration of nuclei far away from the valley of stability 
is a long-term endeavour of nuclear physics with strong relevance for
astrophysical applications. In our present experimental proposal
of a new nuclear reaction scheme, we address the heavy nuclei of the 
$r$-process nucleosynthesis path towards the waiting point at N$=$126, 
where our new production scheme holds promise to bring these extremely 
neutron-rich isotopes into reach of direct experimental studies with
significantly higher yields than accessible with classical radioactive 
ion beam accelerator technology. 
With much more compact high-power, short-pulse laser systems we intend
to develop an optimized production scheme for extremely neutron-rich
fusion products following induced fission from laser-accelerated ion beams.
Exploiting the 'hole-boring' mode of the Radiation Pressure Acceleration 
mechanism will allow 
to generate ion beams of fissile species with solid-state density.
A two-step production scheme of neutron-rich nuclides ('fission-fusion') 
is proposed, where asymmetric fission preceded by a deep inelastic transfer 
reaction will be followed by fusion of the light fission fragments.
Moreover, collective effects reducing the electronic stopping power
in the target are expected for such ultra-dense ion bunches, allowing
to use much thicker targets and thus increasing the fission yield
significantly. The fusion of short-lived, neutron-rich
fission fragment beams with short-lived, neutron-rich fission fragments
in the target will result in very attractive production rates of 
extremely neutron-rich nuclides towards N$=$126 and Z$>$70. Order-of-magnitude
estimates promise fusion rates of several 10$^3$ fusion products per laser pulse,
based on the laser parameters envisaged for the ELI-Nuclear Physics project 
in Bucharest (2 x 150 J, 32 fs). Whereas the present repetition rate
of 1 laser pulse per minute limits the achievable fusion yield, ongoing
development efforts for significantly higher repetition rates (aiming at 
up to 10 Hz) and increased laser energy (aiming at beyond 1 kJ) 
will open the perspective to increase the achievable yields within the
fission-fusion reaction scheme by several orders of magnitude within the
next years.\\
In this way, high-power lasers used for laser ion acceleration can significantly
contribute to access terra incognita in nuclear physics and astrophysical
nucleosynthesis of heavy elements. 

\vspace{10mm}

{\bf Acknowledgement}\\

We acknowledge helpful discussions with M.~Heil, K.L.~Kratz, H.Th.~Janka
and P.~Ring. The authors enjoyed the collaboration with V.~Zamfir, who
is heading the ELI-NP project, opening up many of the new perspectives 
presented here.
We were supported by the DFG Clusters of Excellence:
'Munich Centre for Advanced Photonics' (MAP) and 'Origin and Structure of 
the Universe' (UNIVERSE) and by the DFG Trans\-regio TR18.


\begin{thebibliography}{99}


\bibitem{kratza07}
     K.L.~Kratz, K. Farouqi, B. Pfeiffer, {\it Nuclear physics far from stability 
     and r-process nucleosynthesis}, Prog. in Part. and Nucl. 
     Phys. {\bf 59}, 147 (2007).

\bibitem{arnould07}
     M.~Arnould, S.~Goriely, K.~Takahashi, {\it The r-process of stellar nucleosynthesis: 
     Astrophysics and nuclear physics achievements and mysteries}, 
     Phys. Rep. {\bf 450}, 97 (2007).

\bibitem{janka09}
     I.V~Panov and H.-Th.~Janka, {\it On the dynamics of proto-neutron star 
     winds and r-process nucleosynthesis}, Astr. Astroph. {\bf 494}, 829 (2009).

\bibitem{robinson09}
     A.P.L. Robinson, P. Gibbon, M. Zepf, S. Kar, R.G. Evans and C. Bellei, 
     {\it Relativistically correct hole-boring and ion 
     acceleration by circularly polarized laser pulses},
     Plasma Phys. Control. Fusion {\bf 51} (2009) 024004. 

\bibitem{henig09}
     A. Henig, S. Steinke, M. Schn\"urer, T. Sokollik, R. H\"orlein, D. Kiefer, D. Jung, 
     J. Schreiber, B.M. Hegelich, X.Q. Yan, J. Meyer-ter-Vehn, T. Tajima, P.V. Nickles, 
     W. Sandner, and D. Habs,  
     {\it Radiation-Pressure Acceleration of Ion Beams Driven by Circularly Polarized 
          Laser Pulses},
     Phys. Rev. Lett. {\bf 103}, 245003 (2009).

\bibitem{tajima09} 
    T.~Tajima, D.~Habs, X.~Yan, {\it Laser Acceleration of 
    Ions for Radiation Therapy}, Rev. Accel. Science and Technol. 
    {\bf 2}, 221 (2009).

\bibitem{eli-np}
       http:$\/\/$www.eli-np.ro$\/$

\bibitem{haseltine02}
    E.~Haseltine, http:$\/\/$discovermagazine.com$\/$2002$\/$feb$\/$cover     

\bibitem{rolfs91}
    C.E.~Rolfs, {\it Cauldrons in the Cosmos}, Univ. of Chicago Press (1991).

\bibitem{kratz07}
  K.-L. Kratz, K. Farouqi, B. Pfeiffer, J.W. Truran, C. Sneden 
  and J.J. Cowan, {\it Explorations of the r-Processes: Comparisons between 
  Calculations and Observations of Low-Metallicity Stars},  
  Ap.J. {\bf 662}, 39 (2007).

\bibitem{janka07}
  H.-Th. Janka, K. Langanke, A. Marek, G. Martinez-Pinedo and B. M\"uller,
  {\it Theory of core-collapse supernovae}, 
  Phys. Rep. {\bf 442}, 38 {2007}. 

\bibitem{thielemann04}
  J.J.~Cowan and F.-K. Thielemann, {\it R-Process Nucleosynthesis in Supernovae},
  Physics Today 57 (2004) 47.  

\bibitem{freiburghaus99}
 C. Freiburghaus, S. Rosswog and F.-K. Thielemann,
 {\it r-Process in Neutron Star Mergers},
 Ap. Jour. {\bf 525}, L121 (1999).

\bibitem{janka08}
  R.~Surman, G.C. McLaughlin, M. Ruffert, H.-Th. Janka and W.R. Hix, 
  {\it r-Process Nucleosynthesis in Hot Accretion Disk Flows from Black 
  Hole-Neutron Star Mergers}, Ap. Jour.{\bf679}, L117 (2008).

\bibitem{thielemann05}
  K. Farouqi, C. Freiburghaus, K.-L. Kratz, B. Pfeiffer, T. Rauscher 
  and F.-K. Thielemann,
  {\it Astrophysical conditions for an r-process in the high-entropy wind 
  scenario of type II supernovae}, 
  Nucl. Phys. {\bf A758}, 631c (2005).

\bibitem{pearson96}
    J.M.~Pearson, R. C. Nayak and S. Goriely,
    {\it Nuclear mass formula with Bogolyubov-enhanced shell-quenching: 
    application to r-process}, Phys. Lett. {\bf B387}, 455 (1996).

\bibitem{cowan} 
      J.J. Cowan and C. Sneden,
      {\it Heavy element synthesis in the oldest stars and the early Universe},
      Nature 440, 1151 (2006).

\bibitem{dobaczewki84}
  J.~Dobaczewski, H. Flocard, J. Treiner,
  {\it Hartree-Fock-Bogolyubov description of nuclei near the 
  neutron-drip line}, Nucl. Phys. {\bf A 422}, 103 (1984).

\bibitem{FAIR06}
   FAIR , {\it An International Accelerator Facility for Beams of Ions 
   and Antiprotons}, Baseline Technical Report, GSI (2006),
   http:$\/\/$www.gsi.de$\/$fair$\/$reports$\/$btr.html.

\bibitem{GANIL01}
   SPIRAL II, {\it Detailed Design Study - APD Report}, GANIL (2005),
   http:$\/\/$pro.ganil-spiral2.eu$\/$spiral2$\/$what-is-spiral2$\/$apd

\bibitem{frib}
   http:$\/\/$www.frib.msu.edu$\/$about$\/$msu-frib-proposal

\bibitem{baruah08}
  S. Baruah, G. Audi, K. Blaum, M. Dworschak, S. George, C. Guenaut, 
  U. Hager, F. Herfurth, A. Herlert, A. Kellerbauer, H.-J. Kluge, 
  D. Lunney, H. Schatz, L. Schweikhard and C. Yazidjian,
  {\it Mass Measurements beyond the Major r-Process Waiting Point $^{80}$Zn},
  Phys. Rev. Lett. {\bf 101}, 262501 (2008).

\bibitem{dworschak08}
  M.~Dworschak, G. Audi, K. Blaum, P. Delahaye, S. George, U. Hager, 
  F. Herfurth, A. Herlert, A. Kellerbauer, H.-J. Kluge, D. Lunney, 
  L. Schweikhard and C. Yazidjian, 
  {\it Restoration of the N=82 Shell Gap from Direct Mass Measurements 
  of $^{132,134}$Sn}, Phys Rev. Lett. {\bf 100}, 072501 (2008).

\bibitem{dillmann03}
  I. Dillmann, K.-L. Kratz, A. W\"ohr, O. Arndt, B.A. Brown, P. Hoff, 
  M. Hjorth-Jensen, U. K\"oster, A.N. Ostrowski, B. Pfeiffer, 
  D. Seweryniak, J. Shergur and W. B. Walters,
  {\it N$=$82 Shell Quenching of the Classical r-Process 'Waiting-Point' 
   Nucleus $^{130}$Cd}, Phys. Rev. Lett {\bf 91}, 162503 (2003).

\bibitem{blaum06}
  K.~Blaum, {\it High-accuracy mass spectrometry with stored ions},
  Phys. Rep. {\bf 425}, 1 (2006).

\bibitem{audi95}
   G.~Audi, A.H.~Wapstra,
   {\it The 1995 update to the atomic mass evaluation},
   Nucl. Phys. {\bf A 595}, 409 (1995).

\bibitem{panov05}
  I.V. Panov, E. Kolbe, B. Pfeiffer, T. Rauscher, K.-L. Kratz 
  and F.-K. Thielemann,
  {\it Calculations of fission rates for r-process nucleosynthesis},
  Nucl. Phys. A747 (2005) 633.

\bibitem{ring09}
  P.~Ring, private communication (2010).

\bibitem{moeller97}
   P. M\"oller, J.R. Nix, W.D. Myers and W.J. Swiatecki, 
   {\it Nuclear Ground-State Masses and Deformations},
   Atomic Data Nucl. Data Tabl. {\bf 59}, 185 (1995).

\bibitem{hegelich06}
   B.\,M.~Hegelich, B.J. Albright, J. Cobble, K. Flippo, 
   S. Letzring, M. Paffett, H. Ruhl, J. Schreiber, R.K. Schulze 
   and J.C. Fernandez,
   {\it Laser acceleration of quasi-monoenergetic MeV ion beams},
   Nature {\bf 439}, 441 (2006).

\bibitem{schwoerer06}
   H.~Schw\"orer, S. Pfotenhauer, O. J\"ackel, K.-U. Amthor, B. Liesfeld, 
   W. Ziegler, R. Sauerbrey, K.W.D. Ledingham and T. Esirkepov,
   {\it Laser plasma acceleration of quasi-monoenergetic protons from 
   microstructured targets}, Nature {\bf 439}, 445 (2006).

\bibitem{teravetisyan2006}
   S. ~Ter-Avetisyan, M. Schn\"urer, P.V. Nickles, M. Kalashnikov, 
   E. Risse, T. Sokollik, W. Sandner, A. Andreev and V. Tikhonchuk, 
   {\it Quasimonoenergetic Deuteron Bursts Produced by Ultraintense Laser Pulses},
   Phys.\ Rev.\ Lett. {\bf 96}, 145006 (2006).

\bibitem{snavely00} 
   R.A.~Snavely, M.H. Key, S.P. Hatchett, T.E. Cowan, M. Roth, 
   T.W. Phillips, M.A. Stoyer, E.A. Henry, T.C. Sangster, M.S. Singh, 
   S.C. Wilks, A. MacKinnon, A. Offenberger, D.M. Pennington, K. Yasuike, 
   A.B. Langdon, B.F. Lasinski, J. Johnson, M.D. Perry and E.M. Campbell, 
   {\it Intense High-Energy Proton Beams from Petawatt-Laser Irradiation of Solids},
   Phys.\ Rev.\ Lett.\ {\bf 85}, 2945 (2000).

\bibitem{clarke00} 
   E.~Clarke, K. Krushelnick, J.R. Davies, M. Zepf, M. Tatarakis, F.N. Beg, 
   A. Machacek, P.A. Norreys, M.I.K. Santala, I. Watts, and A.E. Dangor, 
   {\it Measurements of Energetic Proton Transport through Magnetized Plasma 
   from Intense Laser Interactions with Solids},
   Phys.\ Rev.\ Lett. {\bf 84}, 670 (2000).

\bibitem{maksimchuck00} 
   A. Maksimchuck, S. Gu, K. Flippo, D. Umstadter, V.Yu. Bychenkov, 
   {\it Forward Ion Acceleration in Thin Films Driven by a High-Intensity Laser},
   Phys.\ Rev.\ Lett.\ {\bf 84},4108 (2000).

\bibitem{hatchett00} 
   S.\,P.~Hatchett, C.G. Brown, T.E. Cowan, E.A. Henry, J.S. Johnson, M.H. Key, 
   J.A. Koch, A.B. Langdon, B.F. Lasinski, R.W. Lee, A.J. Mackinnon, D.M. Pennington, 
   M.D. Perry, T.W. Phillips, M. Roth, T.C. Sangster, M.S. Singh, R.A. Snavely, 
   M.A. Stoyer, S.C. Wilks and K. Yasuike,
   {\it Electron, photon, and ion beams from the relativistic interaction 
   of Petawatt laser pulses with solid targets},
   Phys.\ Plasma {\bf 7}, 2076 (2000).

\bibitem{robson07}
   L.~Robson, P.T. Simpson, R.J. Clarke, K.W.D. Ledingham, F. Lindau, O. Lundh,
   T. McCanny, P. Mora, D. Neely, C.-G. Wahlstr\"om, M. Zepf, P. McKenna,
   {\it Scaling of proton acceleration driven by petawatt-laser-plasma interactions}
   Nature Physics {\bf 3}, 58 (2007) and refs.\ therein.

\bibitem{macchi05}
   A. Macchi, F. Cattani, T.V. Liseykina and F. Cornolti,
   {\it Laser Acceleration of Ion Bunches at the Front Surface of Overdense 
   Plasmas}, Phys. Rev. Lett. {\bf 94}, 165003 (2005).

\bibitem{klimo08}
   O. Klimo, J. Psikal, J. Limpouch, V.T. Tikhonchuk, 
   {\it Monoenergetic ion beams from ultrathin foils irradiated by 
   ultrahigh-contrast circularly polarized laser pulses},
   Phys. Rev. ST Accl. Beams {\bf 11}, 031301 (2008).

\bibitem{robinson08}
   A.P.L. Robinson, M. Zepf, S. Kar, R.G. Evans and C. Bellei, 
   {\it Radiation pressure acceleration of thin foils with circularly 
   polarized laser pulses}, New J. Phys. {\bf 10}, 013021 (2008).

\bibitem{rykovanov08}
  S.G. Rykovanov, J. Schreiber, J. Meyer-ter-Vehn, C. Bellei, A. Henig, 
  H.C. Wu and M. Geissler, {\it Ion acceleration with ultra-thin foils 
  using elliptically polarized laser pulses}, 
  New J. Phys. {\bf 10}, 113005 (2008).  

\bibitem{robinson09a}
     A.P.L. Robinson, D.-H. Kwon and K. Lancaster, {\it Hole-boring radiation 
     pressure acceleration with two ion species},  
     Plasma Phys. Control. Fusion {\bf 51} (2009) 095006.  

\bibitem{yu10}
     T.-P. Yu, A. Pukhov, G. Shvets, M. Chen \\
     {\it Stable Laser-Driven Proton Beam Acceleration from a Two-Ion-Species
     Ultrathin Foil} \\
     Phys. Rev. Lett. {\bf 105}, 065002 (2010).

\bibitem{chen09}
     M. Chen, A. Pukhov, T.P. Yu, Z.M. Sheng \\
     {\it Enhanced Collimated GeV Monoenergetic Ion Acceleration from a 
          Shaped Foil Target Irradiated by a Circularly Polarized Laser Pulse}\\
     Phys. Rev. Lett. {\bf 103}, 024801 (2009).

\bibitem{yan08}
     X.Q. Yan, C. Lin, Z.M. Sheng, Z.Y. Guo, B.C. Liu, Y.R. Lu, J.X. Fang,
     J.E. Chen \\
     {\it Generating High-Current Monoenergetic Proton Beams by a Circularly
          Polarized Laser Pulse in the Phase-Stable Acceleration Regime} \\
     Phys. Rev. Lett. {\bf 100}, 135003 (2008).

\bibitem{tripathi09}
     V.K. Tripathi, C.S. Liu, X. Shao, B. Eliasson, R.Z. Sagdeev \\
     {\it Laser acceleration of monoenergetic protons in a self-organized
          double layer from thin foil} \\
     Plasma Phys. Control. Fusion {\bf 51} (2009) 024014.

\bibitem{yan09}
     X.Q. Yan, H.C. Wu, Z.M. Sheng, J.E. Chen, J. Meyer-ter-Vehn \\
     {\it Self-Organizing GeV, Nanocoulomb, Collimated Proton Beam
          from Laser Foil Interaction at 7 x 10$^{21}$ W/cm$^2$} \\
     Phys. Rev. Lett. {\bf 103}, 135001 (2009).

\bibitem{yan10}
   X.Q. Yan, T. Tajima, M. Hegelich, L. Yin and D. Habs,
   {\it Theory of laser ion acceleration from a foil target of nanometer 
   thickness}, Appl. Phys. {\bf B 98} 711 (2010).

      
\bibitem{ILE}
   G. Mourou, {\it The APOLLON/ILE 10 PW laser project},
   http://www.nipne.ro/eli$\_$np$\_$workshop/contributions.php

\bibitem{segre77}
   S.~Segr\'e, {\it Nuclei and Particles}, second edt., 
   W.A. Benjamin, RA, London (1977) 

\bibitem{ichimaru73}
   S.~Ichimaru, {\it Basic Principles of Plasma Physics: A Statistical Approach},
   Benjamin, Reading, Ma (1973).

\bibitem{wu09}
     H.\,-C.~Wu {\it et al.}, ``Collective Deceleration''; 
     arXiv:0909.1530v1 [physics.plasm-ph], submitted to Phys. Rev. ST AB.

\bibitem{thirolf02}
     P.G.~Thirolf and D.~Habs, {\it Spectroscopy in the Second and Third 
     Minimum of Actinide Nuclei},
     Prog. in Part. and Nucl. Phys. {\bf 49}, 325 (2002).

\bibitem{metag75}
     V.~Metag., Nukleonica {\bf 20}, 789 (1975).

\bibitem{vandenbosch73}
   R.~Vandenbosch and J.R.~Huizenga, {\it Nuclear fission},
   Academic Press, New York (1973)


\bibitem{pace}
    A. Gavron, {\it Statistical model calculations in heavy ion reactions}, 
    Phys. Rev. {\bf C21} (1980) 230; 
    O.B. Tarasov and D. Bazin, {\it Development of the program LISE: 
    application to fusion-evaporation},
    Nucl. Instr. Meth. {\bf B204} (2003) 174.

\bibitem{quint93}
    A.B.~Quint, W. Reisdorf, K.-H. Schmidt, P. Armbruster, F-P. He{\ss }berger, 
    S. Hofmann, J. Keller, G. M\"unzenberg, H. Stelzer, H.-G. Clerc, 
    W. Morawek, C.-C. Sahm,
    {\it Investigation of the fusion of heavy nearly symmetric systems},
    Z. Phys. {\bf A 346}, 119 (1993).

\bibitem{morawek91}
    W.~Morawek, D. Ackermann, T. Brohm, H.-G. Clerc, U. Gollerthan, 
    E. Hanelt, M. Horz, W. Schwab, B. Voss, K.-H. Schmidt, 
    F.-P. He??berger,
    {\it Breakdown of the compound-nucleus model in the fusion-evaporation 
    process for $^{110}$Pd$+ ^{110}$Pd},
    Z. Phys. {\bf A 341}, 75 (1991).

\bibitem{risac}
   NRC Rare Isotope Science Assessment Committee (RISAC) 
   Report, 2007, National Academies Press, Washington/DC, USA.

\bibitem{kurtu08}
   T. Kurtukian-Nieto, J. Benlliure, K.H. Schmidt, 
   {\it A new analysis method to determine $\beta$-decay half-lives in 
   experiments with complex background},
   Nucl. Instr. Meth. A589 (2008) 472 

\bibitem{plass08}
   W.R. Plass, T. Dickel, U. Czok, H. Geissel, M. Petrick, K. Reinheimer, 
   C. Scheidenberger, M.I.Yavor,
   {\it Isobar separation by time-of-flight mass spectrometry for low-energy 
   radioactive ion beam facilities}, 
   Nucl. Instr. Meth. B266 (2008) 4560.

\bibitem{neumayr07}
   J.B. Neumayr, L. Beck, D. Habs, S. Heinz, J. Szerypo, P.G. Thirolf,
   V. Varentsov, F. Voit, D. Ackermann, D. Beck, M. Block,
   A. Chaudhuri, Z. Di, S.A. Eliseev, H. Geissel, F. Herfurth,
   F.P. He{\ss }berger, S. Hofmann, H.J. Kluge, M. Mukherjee,
   G. M\"unzenberg, M. Petrick, W. Quint, S. Rahaman, C. Rauth,
   D. Rodriguez, C. Scheidenberger, G. Sikler, Z. Wang, C. Weber,
   W.R. Pla{\ss }, M. Breitenfeldt, G. Marx, L. Schweikhard,
   A.F. Dodonov, Y. Novikov, M. Suhonen, {\it The ion catcher device for 
   SHIPTRAP}, Nucl. Inst. Meth. {\bf B244}, 489 (2006).\\
   J.B. Neumayr, P.G. Thirolf, D. Habs, S. Heinz, V.S. Kolhinen,
   M. Sewtz, J. Szerypo, {\it Performance of the MLL-IonCatcher}, 
   Rev. Sci. Instr. {\bf 77}, 065109 (2006). 

\bibitem{block05}
   M. Block, D. Ackermann, D. Beck, K. Blaum, M. Breitenfeld,
   A. Chauduri, A. Doemer, S. Eliseev, D. Habs, S. Heinz, F. Herfurth,
   F.P. Hessberger, S. Hofmann, H. Geissel, H.-J. Kluge, V. Kolhinen,
   G. Marx, J.B. Neumayr, M. Mukherjee, M. Petrick, W. Plass, 
   W. Quint, S. Rahaman, C. Rauth, D. Rodriguez, C. Scheidenberger,
   L. Schweikhardt, M. Suhonen, P.G. Thirolf, Z. Wang, C. Weber,
   and the SHIPTRAP collaboration, {\it The ion trap facility SHIPTRAP},
   Eur. Phys. Jour. A25 (suppl. 1) (2005) 49.

\bibitem{mukh05}
    M. Mukherjee, D. Beck, K. Blaum, G. Bollen, J. Dilling, S. George, 
    F. Herfurth, A. Herlert, A. Kellerbauer, H.-J. Kluge, S. Schwarz, 
    L. Schweikhard, C. Yazidjian,
    {\it ISOLTRAP: An on-line Penning trap for mass spectrometry on short-lived 
    nuclides}, Eur. Phys. J. A 35, 1-29 (2008).
 
\bibitem{block10}
    M. Block, D. Ackermann, K. Blaum, C. Droese, M. Dworschak,
    S. Eliseev, T. Fleckenstein, E. Haettner, F. Herfurth,
    F.P. He{\ss }berger, S. Hofmann, J. Ketelaer, J. Ketter,
    H.-J. Kluge, G. Marx, M. Mazzocco, Yu.N. Novikov, W. R. Pla{\ss },
    A. Popeko, S. Rahaman, D. Rodriguez, C. Scheidenberger,
    L. Schweikhard, P.G. Thirolf, G.K. Vorobyev, C. Weber,
    {\it Direct mass measurements above uranium bridge the gap to the 
    island of stability},
    Nature {\bf 463}, 785 (2010).


   
\end{thebibliography}
\end{document}